\documentstyle[11pt]{report}
\begin{document}
\def\mtxt#1{\quad\hbox{#1}\quad}
\def\es{\!=\!}
\def\ov{\over}
\def\pa{\partial}
\def\pamu{{\partial_\mu}}
\def\panu{{\partial_\nu}}
\def\al{\alpha}
\def\be{\beta}
\def\si{\sigma}
\def\gam{\gamma}
\def\pr{\prime}
\def\lam{\lambda}
\def\tr{\hbox{tr}\,}
\def\ta{{\tilde a}}
\def\tb{{\tilde b}}
\def\tc{{\tilde c}}
\def\tal{{\tilde\alpha}}
\def\tsi{{\tilde\sigma}}
\def\tbe{{\tilde\beta}}
\def\tga{{\tilde\gamma}}
\def\tde{{\tilde\delta}}
\def\ti{{\,\tilde i}}
\def\tj{{\,\tilde j}}
\def\hnn{\hat{0}}
\def\hr{\hat{R}}
\def\hi{\hat{I}}
\def\jm{\hat{N}}
\def\hio{\hat{I}_{\xi_1\lambda_1}}
\def\hit{\hat{I}_{\xi_2\lambda_2}}
\def\dxila{\delta_{\xi,\lambda}}
\def\dq{\dot{q}}
\def\cm{{\cal M}}
\def\cn{{\cal N}}
\def\ch{{\cal H}}
\def\cl{{\cal L}}
\def\ctal{{C_\tal}}
\def\ctbe{{C_\tbe}}
\def\ctga{C_\tga}
\def\tttr{t^\tga_{\tal \tbe}}
\def\tttw{t^\tbe_\tal}
\def\str{f^c_{ab}}
\def\ha{{1\over 2}}
\def\una{{\vec A}}
\def\unb{{\vec B}}
\def\und{{\vec D}}
\def\unpi{{\vec \pi}}
\def\unpa{{\vec\pa}}
\title{On the Symmetries of Hamiltonian Systems}
\author{V. Mukhanov\thanks{On leave of absence from Institute for
Nuclear Research, Moscow 117312, Russia}and A. Wipf\\
Institut f\"ur Theoretische Physik\\
Eidgen\"ossische Technische Hochschule\\
H\"onggerberg, CH-8093 Z\"urich, Switzerland}
\date{ETH-TH/94-04, January 1994}
\maketitle
\begin{abstract}
In this paper we show how the well-known local symmetries of Lagrangean
systems, and in particular the diffeomorphism invariance, emerge
in the Hamiltonian formulation. We show that only the constraints
which are linear in the momenta generate transformations which correspond
to symmetries of the corresponding Lagrangean system. The nonlinear
constraints (which we have, for instance, in gravity,
supergravity and string theory) rather generate the dynamics of the
corresponding Lagrangean system. Only in a very special combination
with "trivial" transformations proportional to the equations of motion
do they lead to symmetry transformations. We reveal the importance
of these special "trivial" transformations for the interconnection
theorems which relate the symmetries of a system with its dynamics.
We prove these theorems for general Hamiltonian systems. We
apply the developed formalism to concrete physically relevant systems
and in particular those which are diffeomorphism invariant.
The connection between the parameters of the symmetry transformations
in the Hamiltonian- and Lagrangean formalisms is found.
The possible applications of our results are discussed.
\end{abstract}
\newcommand{\refb}[1]{(\ref{#1})}
\newcommand{\eqngr}[2]{\par\parbox{11cm}
{\begin{eqnarray*}#1\\#2\end{eqnarray*}}\hfill
\parbox{1cm}{\begin{eqnarray}\end{eqnarray}}}

\newcommand{\eqngrr}[3]{\par\parbox{11cm}
{\begin{eqnarray*}#1\\#2\\#3\end{eqnarray*}}\hfill
\parbox{1cm}{\begin{eqnarray}\end{eqnarray}}}

\newcommand{\eqngrl}[3]{\par\parbox{11cm}
{\begin{eqnarray*}#1\\#2\end{eqnarray*}}\hfill
\parbox{1cm}{\begin{eqnarray}\label{#3}\end{eqnarray}}}

\newcommand{\eqngrrl}[4]{\par\parbox{11cm}
{\begin{eqnarray*}#1\\#2\\#3\end{eqnarray*}}\hfill
\parbox{1cm}{\begin{eqnarray}\label{#4}\end{eqnarray}}}
\chapter{Introduction}
Local symmetries play a very important role in all field
theories being relevant in physics. The actions of
such theories are invariant with respect to some group of local
transformations. For example, for Yang-Mills theories these are
the gauge transformations, for string theory and gravity
diffeomorphisms and for supersymmetric
theories coupled to gravity local
supersymmetry transformations. These symmetries are quite transparent
in the Lagrangean formulation and this is seen as one
of the main virtues of this approach. Actually the Lagrangean
of a theory is constructed such that it is invariant under
gauge transformations and/or diffeomorphisms.

If we go from the Lagrangean to the first order Hamiltonian formalism
then at first glance it seems that these symmetries are not
manifest. This applies especially to diffeomorphism invariant
theories and is of much relevance in general relativity
\cite{bk72,ik85,lw90,hh91}. One of the purposes of this paper is to
show that one can construct the symmetries
of constrained Hamiltonian systems in an explicit manner.

It was found in \cite{t77,fv77} that the first order action is invariant
with respect to {\it infinitesimal} time-dependent transformations
generated by the first class constraints if the Lagrangean
multipliers are simultaneously transformed. However, we shall
see that these transformations correspond to Lagrangean symmetries
only if the constraints are {\it linear in momenta}. For instance,
this is the case for Yang-Mills theories,
where all gauge transformations (including time-dependent ones) can
be recovered in such a manner in the Hamiltonian formalism.

For the constraints which are nonlinear in the momenta (as they
exist in diffeomorphism invariant theories, e.g.
gravity or string theory) this is not true anymore.
The nonlinear constraints by themselves
{\it do not} generate transformations which correspond
to symmetries of the corresponding Lagrangean system.
They are rather responsible for the dynamics of such
systems.

Although the transformations generated by the nonlinear constraints
are still symmetries of the Hamiltonian system (which
cannot be identified with Lagrangean symmetries) it is not clear
whether they are of any relevance, since only their infinitesimal
form is known. It is not obvious whether for nontrivial theories
they can be exponentiated, that is can be iterated to finite transformations.

The action in the Hamiltonian (and even Lagrangean) formalism is
also invariant with respect to so-called {\it infinitesimal}
'trivial' transformations \cite{vt82,h89,ht92} which are proportional to
antisymmetric combinations of the equations of motion and do not vanish off
mass-shell. This huge class of additional transformations exists even in
theories without local symmetries. It is clear that
most of them (or sometimes even all) are irrelevant and
can safely be ignored \cite{vt82,h89,ht92}. However, we shall
see that not all of the "trivial" transformations are really
unimportant for the systems with nonlinear constraints. Indeed,
we shall demonstrate that all Lagrangean symmetries can be recovered
in the Hamiltonian formalism only if we consider the transformations
generated by the nonlinear constraints in a very special combination
with {\it particular} "trivial" transformations. The combined
transformations can be exponentiated since they correspond to
known Lagrangean symmetries. Thus, {\it not all} of the trivial
transformations are irrelevant for systems with nonlinear constraints,
although they may be ignored for particular perturbative questions
\cite{ht92}. However, this is not alway the case. In particular
we shall see later that it is impossible to
get the theorems which relate the dynamics of a super-hamiltonian
system with its symmetry properties (e.g. the interconnection theorems
in general relativity) if we neglect the trivial transformations.
Also, when one ignores them this can lead to wrong results in
nonperturbative calculations. One last remark concerns the identification
of transformations generated by the constraints with the
Lagrangean symmetries on mass-shell. It seems that this
identifictaion of {\it infinitesimal} transformations is meaningless,
since on mass-shell any infinitesimal transformation can be
viewed as "symmetry transformation" since solutions of the equations
of motion are stationary points of the action.

The questions which we address in this paper are
the following. First we investigate how one recovers
and generalizes the local Lagrangean symmetries in the first
order Hamiltonian formalism. This question has also been raised
recently in \cite{htz90}. However, our approach is very different and
can be viewed as complimentary to that in \cite{htz90}. Also we explicitly
reveal the connection between the parameters of the transformations
in the Hamiltonian and Lagrangean formalisms in most physically
important theories. Some of these results (but not all) can be found
in the literature and our purpose here will be to clarify the confusing
points which still exist. The other question concerns the
difference between linear and nonlinear in momenta constraints.
We will show that the transformations generated by the nonlinear
constraints always take any trajectory which belongs to the subspace
where the Lagrangean system lives \footnote{The subspace on which
the momenta and velocities are related by the first half of Hamilton's
equations} away from this subspace. Hence these transformations
cannot correspond to Lagrangean symmetries. The role of the "trivial"
transformations is to project the trajectory back to this subspace.
The nonlinear constraints themselves rather generate the
dynamics of the corresponding Lagrangean systems.

We will follow in detail how the closed Lie algebra belonging to the
diffeomorphism group arises in a natural manner in the Hamiltonian
formalism. We clarify the connection between the symmetry properties
of the system and its dynamics and prove the so-called
"interconnection" theorem \cite{i92} for general constrained Hamiltonian
systems entirely in the Hamiltonian formalism.
This theorem plays a crucial role in the Dirac quantization program
and also in the Hamilton-Jacobi appraoach to classical general relativity.
It will be shown that this theorem is nontrivial only for theories
with an infinite number of degrees of freedom and only if there are
nonlinear in momenta constraints. The special role played by
the trivial transformations in proving it is emphasized.
Most of our considerations are classical
and we comment on the corresponding problems in the quantized
theories at the end of the paper.
\par
The paper is organized as follows. In the second section we describe the
symmetries of general first order Hamiltonian systems. In the
subsequent sections we apply the results to gauge theories,
the relativistic particle, the locally supersymmetric relativistic
particle, bosonic string and to general relativity. We show that the
local symmetries of Hamiltonian systems coincide with the
local symmetries of the corresponding Lagrangean systems by
revealing the connection between the parameters of the corresponding
groups for the Hamiltonian and Lagrangean systems. In the last
chapter we discuss why from our point of view the Hamiltonian
formalism is more 'fundamental' than the Lagrangean one,
in particular for the quantized theories, and describe the
possible applications of the developed formalism.
\chapter{Symmetry Transformations}
We shall consider a general first order Hamiltonian system with
constraints, the action of which is
\begin{equation}
S=\int \Big(p^{\ti}\dq_\ti-\cn^\tal C_\tal (p,q)-\ch(p,q)\Big)dt.\label{mw1}
\end{equation}
If the system contains fermions then
some of the variables $p,q,\cn$ will be of Grassmannian type.
The first order action \refb{mw1} describes both systems with a finite
or infinite number of degrees of freedom since the following
condensed notation \cite{dw65} is assumed: the indices $\ti,\tal$ are
supposed to be composite ones, that is they may contain discrete
and continuous variables.
For systems with a finite number of degrees of freedom
$\tal =\al$ and $\ti =i$ are discrete.
For field theories $\ti =\{i,x\} $ and $\tal =\{\al,x\}$, where $i$ and $\al$
are some discrete (internal) indices and in $d$ space-time
dimensions $x=\{x^1,\dots,x^{d-1}\}$ are the $d\!-\!1$ space coordinates.
For example, for a scalar field $q_\ti (t)=\phi_x(t)=\phi(x,t)$
and for a vector field $q_\ti(t)=A_{i,x}(t)=A_i(x,t)$, where
in $4$ spacetime dimensions $i=1,2,3$. We adopt the Einstein
convention for repeated  indices in 'up' and 'down' positions, that
is we assume summation over discrete repeated indices and integration
over continuous ones, for example
\begin{equation}
\xi^x p^{i,x}\dq_{i,x}=\sum_i\int dx \xi(x)p^i(x)\dq_i(x),
\end{equation}
but
\begin{equation}
\xi^x p^{i,x} q_i^x=\sum_i \xi(x)p^i(x)q_i(x),\qquad\hbox{no integration.}
\end{equation}
Also, we shall not distinguish $q_{i,x}$ and $q_i^x$ and use the
position of the continuous index just to indicate when we should
integrate. Sometimes it will be convenient to resolve the
composite index $\ti$ (or $\tal$) as $i,x$ (or $\al ,x$).
The dot always denotes derivative with
respect to time $t$ on which $p,q$ and $\cn$ may depend.

For first class Hamiltonian system \cite{d64} the contraints
$C_\tal $ and Hamiltonian $\ch$ form a closed algebra with respect
to the standard Poisson bracket $\{.,.\}$ (possibly
extended to fermionic variables, in which case the algebra
is graded \cite{c76}):
\begin{equation}
\{\ctal ,\ctbe\}=\tttr \ctga\qquad \hbox{and}\qquad
\{\ch,\ctal\}=\tttw \ctbe,\label{mw2}
\end{equation}
where  the $t$'s are the structure coefficients
\footnote{If some constraints depend explicitly on time, then
we should add $\pa_t\ctal$ to the right hand side of the
second relation.}. These
coefficients may depend on the canonical variables. For field theories
$\tttr\equiv t^{\gamma z}_{\alpha x\,\beta y}$ and $\tttw
\equiv t^{\beta y}_{\al x}$.

The equations of motion resulting from the variation of the
action \refb{mw1} with respect to $q,p$ and the Lagrangean multipliers
$\cn$
\begin{equation}
\delta S=\int\Big(\delta p^\ti EM(q_\ti)-\delta q_\ti EM(p^\ti)
-\delta \cn^\tal C_\tal\Big) dt +\hbox{bound. terms}\label{mw3}
\end{equation}
are
\eqngrrl{EM(q_\ti)&\equiv& \dot{q}_\ti-\{q_\ti,\cn^\tbe C_\tbe+\ch\}=0,}
{EM(p^\ti)&\equiv& \dot{p}^\ti-\{p^\ti,\cn^\tbe C_\tbe+\ch\}=0,}
{C_\tal&=&0.}{mw4}
Below we shall often use these abbreviations $EM(q)$ and $EM(p)$
for the left hand sides in \refb{mw4}. Of course, on mass shell we have
$EM=0$, but off mass shell either $EM(q)$ or $EM(p)$ (or both)
does not vanish.

To go from the Hamiltonian to the Lagrangean formalism we should
express the momenta in terms of the velocities
via the Hamiltonian equations $EM(q_\ti)\es 0$. Thus not all
off mass-shell trajectories of the Hamiltonian system can be
considered in the Lagrangean formalism, but only those for which
$EM(q)\es 0$. Hence one can say that the Lagrangean
system lives only in the subspace $\cm$ of the 'extended phase space'
defined by the conditions
\begin{equation}
\cm:EM(q_\ti)\equiv \dot{q}_\ti -\{q_\ti,\cn^\tbe C_\tbe+\ch\}=0\label{mw5}
\end{equation}
Clearly the space of trajectories in phase space where the Hamiltonian
system lives is much bigger than the space of Lagrangean trajectories.

The action \refb{mw1} is invariant (up to boundary terms) with respect to
the infinitesimal transformations generated by the constraints
if the Lagrangean multipliers are transformed simultaneously
\cite{t77,fv77}:
\eqngrrl
{\delta_\lam q^x_i&=&\{q_i^x,\lam^\tbe C_\tbe\},}
{\delta_\lam p^{ix}&=&\{p^{ix},\lam^\tbe C_\tbe\},}
{\delta_\lam \cn^\tal &=&\dot{\lam}^\tal-\lam^\tbe\cn^\tga t^\tal_{\tga\tbe}
-\lam^\tbe t^\tal_\tbe.}{mw6}
The parameters $\lam^\tal=\lam^\al(\cn,x,t)$ in \refb{mw6}
are the parameters of the infinitesimal transformations.
The order in which $\lam$ enters
in \refb{mw6} is important if some of the variables are of Grassmannian type.
We shall only consider the case when the parameters $\lam$
depend explicitly on spacetime coordinates and Lagrangean multipliers,
since this suffices to cover all known physically relevant theories\footnote{
In principle, we could consider more general transformations
for which the $\lam$ would also depend on the canonical
variables $q$ and $p$. One can show that in this case the action
\refb{mw1} is also invariant with respect to infinitesimal transformations
generated by the constraints if $\cn^\tal$ are transformed as
$\delta_\lam \cn^\tal=\pa_t\lam^\tal(q,p,t)
-\lam^\tbe\cn^\tga t^\tal_{\tga\tbe}-\lam^\tbe t^\tal_\tbe
-\cn^\tbe\{C_\tbe,\lam^\tal\}-\{\ch,\lam^\tal\}.$}.
Because of this $\cn$-dependence
we should keep $\lam$ inside the Poisson bracket
even for purely bosonic theories since if we calculate the
commutator of two subsequent infinitesimal transformations,
then the parameter $\lam$ of the second transformation will depend on $q,p$
if the structure coefficients depend on the canonical variables.
It is not difficult to see that the variation of the action
\refb{mw1} under these transformations leads only to the boundary terms
\begin{equation}
\delta_\lam S=\Big(p^\ti\delta_\lam\;q_\ti-\lam^\tal C_\tal\Big)
\vert_{t_i}^{t_f}.
\end{equation}
This term can be removed even if the parameters $\lam$ do not vanish
at the boundaries if we add to the action \refb{mw1} the total
derivative of some function $Q(p,q)$ which satisfies the
equation
\begin{equation}
{\delta Q\over \delta q_\ti}\delta_\lam q_\ti +
{\delta Q\over \delta p^\ti}\delta_\lam p^\ti =\lam^\tal C_\tal-
p^\ti\delta_\lam q_\ti.
\end{equation}
The question which naturally arise here is the following:
do the symmetry transformations \refb{mw6} correspond to Lagrangean
symmetries, that is are they, for instance, the diffeomorphism
transformations in general relativity and string theory?

As we shall see below the answer to this question crucially
depends on the momenta dependence of the constraints.
If the constraints are linear in the momenta, then the answer is yes.
However, it is not the case if some of the constraints are nonlinear.
The reason is that the transformation \refb{mw6}
generated by a nonlinear constraint take a trajectory in $\cm$
(see \refb{mw5}) away from it and the transformed trajectory can not be
viewed as a trajectory of the Lagrangean system, since $EM(q)\es 0$
does not hold anymore. To proceed in this case we should consider
extra compensating symmetry transformations of the Hamiltonian system.

Actually the set of infinitesimal off mass-shell transformations
which leave the action \refb{mw1} invariant is much bigger than \refb{mw6}.
Any infinitesimal transformation $(\delta q,\delta p,\delta \cn)$
orthogonal to the (functional) gradient $\nabla S=(-EM(p),EM(q),-C)$
leaves the action invariant, as can be easily seen from \refb{mw3}.
Hence we could add to the transformations generated by the constraints
for example any transformation of the form
\eqngrrl
{\delta q_\ti &=&EM(q_\tj )\xi^\tj_{\;\,\ti} + EM(p^\tj )\eta_{\tj\ti },}
{\delta p^\ti &=&EM(p^\tj )\xi^\ti_{\;\,\tj} + EM(q_\tj )\zeta^{\tj\ti },}
{\delta \cn^\tal &=&0,}{mw7}
where $\xi^\tj_{\;\,\ti}$ are arbitrary 'matrices' (kernels) and the
$\eta_{\tj\ti},\;\zeta^{\tj\ti}$ are antisymmetric.
Generically such transformations are nonlocal, and they exist
for all systems even for those without any symmetries.

Most of them are actually irrelevant for the physically interesting
transformations \cite{vt82,h89,ht92}. However, if some of the constraints
are nonlinear then, as we will see, the  particular "trivial" transformations
from \refb{mw7} play an important role for recovering the Lagrangean
symmetries in the Hamiltonian formalism.

We will show that in all theories containing only one nonlinear
constraint (e.g. gravity and string theory) we need only very special
transformations from \refb{mw7}, namely
\begin{equation}
\delta_\xi q_i^x=EM(q^x_i)\xi^x\quad\hbox{and}\quad
\delta_\xi p^{ix}=EM(p^{ix})\xi^x\label{mw8}
\end{equation}
to recover all Lagrangean symmetries.

Only for theories with several nonlinear constraints
do we need extra transformations from \refb{mw7} in addition to \refb{mw6}
to recover all off mass-shell Lagrangean symmetries
in the Hamiltonian approach.
For a system with only one nonlinear constraint we consider the
combined transformations
\begin{equation}
\hi_{\xi,\lam}\,F(q,p,\cn )=F\big(\hi_{\xi,\lam}\;q,
\hi_{\xi,\lam}\;p,\hi_{\xi,\lam}\;\cn\big),\qquad
\hi_{\xi,\lam}=\hat{1}+\dxila+\cdots,\label{mw9}
\end{equation}
where
\eqngrrl
{\dxila\, q^x_i&=&EM(q^x_i)\xi^x + \{q_i^x,\lam^\tbe C_\tbe\},}
{\dxila\, p^{ix}&=&EM(p^{ix})\xi^x+\{p^{ix},\lam^\tbe C_\tbe\},}
{\dxila\, \cn^\tal &=&\dot{\lam}^\tal-\lam^\tbe\cn^\tga t^\tal_{\tga\tbe}
-\lam^\tbe t^\tal_\tbe.}{mw10}
The number of functions $(\xi,\lam^\al)$ which appear here is
equal to the number of constraints (per point of space) plus one.
This seems strange since for all constrained theories
the number of parameters in the Lagrangean symmetry transformations
is equal to the number of constraints. Thus not all of the parameters
in \refb{mw10} should be independent for these transformations to be
symmetries of the corresponding Lagrangean systems.
To understand why we need the "trivial" transformations \refb{mw8}
in additions to \refb{mw6} and to reveal the connections between the
parameters $\xi$ and $\lam^\al$ we derive the conditions under which the
transformations \refb{mw10} can be viewed as Lagrangean symmetries.

For that the transformations \refb{mw10} should at least leave the
subspace $\cm$ (see \refb{mw5}) in which the Lagrangean
system lives, invariant. That is, they should leave any
trajectory which belongs to the subspace $\cm$
in this subspace. The necessary conditions for that can
be gotten by varying \refb{mw5} as follows
\eqngrl
{{d\ov dt}\big(\dxila q_\ti\big) &=&
{\delta^2(\ch+\cn^\tsi C_\tsi)\over \delta p^\ti\delta q_\tj}\;\dxila q_\tj}
{&+&{\delta^2(\ch+\cn^\tsi C_\tsi)\over \delta p^\ti\delta p^\tj}\;\dxila p^\tj
+\{q_\ti,\dxila \cn^\tsi C_\tsi\}.}{mw11}
Thus the transformations $\delta q,\;\delta p$ and $\delta\cn$ should
satisfy this equation in the subspace $\cm$.
If this is not the case, then the trajectories for which
\begin{equation}
EM(q)=0\Longleftrightarrow p^\ti =f^\ti (\dot q^\tj,q^\tj,\cn^\tal)\label{mw12}
\end{equation}
are transformed into trajectories for which this equality fails
and they cannot be viewed as trajectories of the corresponding
Lagrangean system. The transformation of $p$ which would follow from
\refb{mw12} as
\begin{equation}
\delta p^\ti={\pa f^\ti\ov\pa q_\tj}\delta q_\tj+{\pa f^\ti\ov
\pa \dot q_\tj}\delta \big(\dot q_\tj\big)+{\pa f^\ti\ov \pa \cn^\tal}
\delta\cn^\tal\label{mw13}
\end{equation}
would be different from the transformation \refb{mw10} for $p$. Thus
\refb{mw10}
will not be a Lagrangean symmetry for which \refb{mw13} must hold.
Substituting \refb{mw10} into \refb{mw11} this condition simplifies to
\begin{equation}
{\delta^2(\ch+\cn^\tsi C_\tsi)\over \delta p^{ix}\delta p^{jy}}EM(p^{jy})\xi^y=
{\delta^2 C_\tsi\over \delta p^{ix}\delta p^{jy}}EM(p^{jy})\lam^\tsi.
\label{mw14}\end{equation}
The point is that this equation relates $\xi$ and $\lam^\al$ and if it
holds then the phase space transformations \refb{mw10} can be
interpreted as Lagrangean symmetries. At the same time the number of free
functions becomes equal to the number of constraints as it should be.

Let us note that the "trivial" transformations \refb{mw8} alone
do not satisfy \refb{mw14} for off mass-shell trajectories
if the Hamiltonian $\ch$ and/or $C_\tal$ are nonlinear in momenta. Hence
they cannot be identified with the Lagrangean symmetries.

If some constraints $C_\tal$ are nonlinear, then the transformations \refb{mw6}
generated by them alone also cannot satisfy \refb{mw14} for
off mass-shell trajectories. Hence they cannot be viewed as Lagrangean
symmetries either. Only when the transformations generated by the
nonlinear constratins are taken in a very special
combination with the "trivial" transformations \refb{mw8} one can
satisfy the condition \refb{mw14}. The reason why the nonlinear constraints
alone do not generate the Lagrangean symmetries is simple.
They always take off mass-shell trajectories away form the subspace
$\cm$, where the Lagrangean system lives. The "trivial"
transformations \refb{mw8} return the trajectories back in $\cm$, if
for the given $\lam^\al$ in \refb{mw10} we take the appropriate $\xi(\lam^\al)$
to satisfy \refb{mw14}. They play the role of compensating transformations.
As we shall see later, the nonlinear constraints themselves generate
the dynamics for Lagrangean systems in the subspace $\cm$.

Now we would like to consider two important examples:
\paragraph{Gauge Invariance.} If the constraints $C_\tal$ are linear
and $\ch$ at least quadratic in the momenta then only for $\xi^z=0$
can equation \refb{mw14} be satisfied
\footnote{If $\ch$ and all constraints are linear in momenta
then the Hamiltonian system is strongly degenerate.}.
So, in this case the transformations which are generated by
the constraints will also be symmetry transformations for the
corresponding Lagrangean system. We shall call them {\em gauge
transformations}:
\begin{equation}
\hat{G}_\lam =\hi _{\xi=0, \lam}\Longrightarrow
\Big\{{\delta_\lam q_\ti=\{q_\ti,\lam^\tbe C_\tbe\},\quad
\delta_\lam p^\ti=\{p^\ti,\lam^\tbe C_\tbe\},\atop
\delta_\lam\cn^\tal =\dot{\lam}^\tal-\lam^\tbe\cn^\tga t^\tal_{\tga\tbe}
-\lam^\tbe t^\tal_\tbe.}\label{mw15}
\end{equation}
For example, in Yang-Mills theories all constraints are linear
in the momenta and (as we shall see in the next section) the
finite gauge transformations can be recovered as transformations
generated only by the constraints ($\xi=0$) in the Hamiltonian formalism
\footnote{Another interesting class of theories where all constraints
are linear in momenta are the constrained Wess-Zumino-Novikov-Witten
theories \cite{frr92}.}. Thus the extra transformations \refb{mw8} are
irrelevan
t
in this case.
\paragraph{Reparametrization invariance.} Very often the reparametrization
invariance of a Lagrangean system, if it exists, is identified with the
gauge invariance \refb{mw15} in the Hamiltonian formalism.
As we shall see they are actually very different.

If some of the constraints in \refb{mw1} are nonlinear then
it is obvious that the transformations generated by the constraints
only ($\xi=0$) do not satisfy \refb{mw14}.
However, in all known theories with nonlinear constraints
$\ch=0$ and the condition \refb{mw14}
can be satisfied if we impose some functional
dependence between $\lam$ and $\xi$ in \refb{mw10} so that $\xi\neq 0$
for such theories. Thus the nonlinear constraints generate the
Lagrangean symmetries only in very special combination with 'trivial'
transformations from \refb{mw8}. More explicitly taking $\lam^\tsi$ to be
$\lam^{\si z}=\cn^{\si z}\xi^z$ in \refb{mw14} we reduce this equation to
\begin{equation}
\cn^{\si z}(\xi^y-\xi^z)
{\delta^2 C_{\si z}\over \delta p^{ix}\delta p^{jy}}
EM(p^{jy})=0.
\end{equation}
One sees at once that if
\begin{equation}
{\delta^2 C_{\si z}\over \delta p^{ix}\delta p^{jy}}\sim \delta(z-y)
\end{equation}
then even for constraints nonlinear in the momenta the equation
\refb{mw14} is satisfied off mass shell ($EM(p)\neq 0$). From that
it follows immediately that the transformations \refb{mw10}
with $\lam^{\si z}=\cn^{\si z}\xi^z$ are symmetry transformations
for the corresponding Lagrangean system if $\ch=0$.
We shall call this symmetry {\em reparametrization invariance}:
$\hr_\xi = \hi_{\xi,\lam^{\si z}=\cn^{\si z}\xi^z}$. The explicit
form of the reparametrization transformations generalized to field
theories is
\eqngrrl
{\delta_\xi q_i^x&=&\dot{q}^x_i\xi^x+(\xi^y-\xi^x)\{q^x_i,\cn^{\be y}C_{\be
y}\}
}
{\delta_\xi p^{ix}&=&\dot{p}^{ix}\xi^x+(\xi^y-\xi^x)\{p^{ix},\cn^{\be y}C_{\be
y
}\}}
{\delta_\xi \cn^{\al x}&=&(\cn^{\al x}\xi^x)^\cdot-\xi^y\cn^{\beta y}
\cn^{\gamma z}t^{\al x}_{\gamma z,\beta y}.}{mw16}
We would like to remind that according to our notation we assume here
integration over $y$ and $z$ but no integration over $x$.
For systems with a finite number of degrees of freedom the
second terms or the right hand sides are absent and \refb{mw16}
has a familiar form.

If several constraints are nonlinear in momenta then there are
extra reparametrisation transformations in addition to \refb{mw16}.
They can be obtained by combining the transformations generated
by the constraints with 'trivial' transformations \refb{mw7} in
such a manner that \refb{mw14} is fulfilled (see section 5).

In some of the theories we shall study (string, gravity)
there are both linear and nonlinear constraints.
For such theories the symmetry transformations
which correspond to Lagrangean symmetries are combinations
of gauge transformations (generated by the linear constraints) and
reparametrization transformations.
\paragraph{Algebra of transformations}
To construct the finite transformations we need to apply
the infinitesimal transformations many times. To be successful
in this 'exponentiation' it is clear that the following
{\em necessary} condition should be fulfilled: The algebra of infinitesimal
transformations should be closed, that is the commutator of two
subsequent transformations should be a transformation of the same
type. To check the algebra of
transformations let us calculate the result for the commutator of two
subsequent infinitesimal transformations \refb{mw10} with
parameters $\xi_1,\lam_1$ and $\xi_2,\lam_2$ respectively.
For an arbitrary algebraic function $F(q,p)$ of the canonical
variables (for example $F\!=\!q$ or $F\!=\!p$) a rather lengthy but
straightforward calculation yields the commutator
\begin{eqnarray}
&&\big[\hit ,\hio\big]\,F^x(q,p)=
\Big({\delta F^x\over \delta q_i^z}EM(q_i^z)+(q\to p)\Big)
(\dot{\xi}_1^z\xi^z_2-\xi_1^z\dot{\xi}_2^z)\nonumber\\
&&\quad
+\Big((\xi_2^x-\xi_2^y)\lam_1^\tga -
\cn^\tga \xi_2^x\xi_1^y -(1\leftrightarrow 2)\Big)
\Big(\big\{F^x,{\delta C_\tga\over \delta q^y_j}\big\}EM(q_j^y)+(q\to p)\Big)
\nonumber\\
&&\quad
-(\xi_2^x\xi_1^y-\xi_1^x\xi_2^y)\Big(\big\{F^x,{\delta \ch\over \delta
q_j^y}\big\}
EM(q_j^y)+(q\to p)\Big)+\{F^x,\bar{\lam}^\tga C_\tga\}\label{mw17}
\end{eqnarray}
and correspondingly for the Lagrangean multipliers one has
\eqngrl{
&&\big[\hit ,\hio]\cn^\tal=
\big(\hi_{\bar{\lam}}-1\big)\cn^\tal+\lam_2^\tde\lam_1^\tga\Big(
\dot{t}^\tal_{\tga\tde}-\{t^\tal_{\tga\tde},\cn^\tsi C_\tsi +\ch\}\Big)}
{&&\quad -(\lam_2^\tga\xi_1^x-\lam_1^\tga\xi_2^x)\Big(
{\delta\over \delta q_i^x}(\cn^\tbe t^\tal_{\tbe\tga}+t^\tal_\tga)EM(q^x_i)+
(q\to p)\Big),}{mw18}
where we have introduced
\begin{equation}
\bar{\lam}^\tal=\lam_1^\tsi\lam_2^\tbe t^\tal_{\tbe\tsi}
+{\delta \lam_2^\tal\over \delta \cn^\tbe}\delta_{\lam_1}\cn^\tbe
-{\delta \lam_1^\tal\over \delta \cn^\tbe}\delta_{\lam_2}\cn^\tbe.\label{mw19}
\end{equation}
In deriving \refb{mw17} and \refb{mw18} we used the identities
\begin{equation}
\big(\lam_1^\tga\lam_2^\tde-\lam_2^\tga\lam_1^\tde\big)\big(
\{t^\tal_{\tsi\tga},C_\tde\}+t^\tbe_{\tsi\tga}t^\tal_{\tbe\tde}\big)
=\lam_1^\tga\lam_2^\tde\big(t^\tbe_{\tga\tde}t^\tal_{\tsi\tbe}
-\{t^\tal_{\tga\tde},C_\tsi\}\big),
\end{equation}
and
\begin{equation}
\big(\lam_1^\tga\lam_2^\tde-\lam_2^\tga\lam_1^\tde\big)\big(
\{t^\tal_\tga,C_\tde\}+t^\tbe_\tga t^\tal_{\tbe\tde}\big)
=\lam_1^\tga\lam_2^\tde\big(t^\tbe_{\tga\tde}t^\tal_\tbe
-\{t^\tal_{\tga\tde},\ch\}\big),
\end{equation}
which follow from the Jacobi identities for
$\{\{C_\tsi,\lam_1^\tga C_\tga\},\lam_2^\tsi C_\tsi\}$ and
for \linebreak
$\{\{\ch,\lam_1^\tga C_\tga\},\lam_2^\tsi C_\tsi\}$.\footnote{
With the exception of section 5 we consider for simplicity only the bosonic
case from now on.}
Also we took into account that if the variables $q,p$ are transformed
to new variables $\tilde{q}=q+\triangle q$ and
$\tilde{p}=p+\triangle p$, then the Poisson bracket of some
quantities
$A(\tilde{q},\tilde{p})$ and $B(\tilde{q},\tilde{p})$
with respect to $\tilde{q},\tilde{p}$ are related with the Poisson
brackets of $A(q,p)$ and $B(q,p)$ with respect to the old variables
in first order in $\triangle q,\;\triangle p$
in the following manner
\eqngr{
\{A(\tilde{q},\tilde{p}),B(\tilde{q},\tilde{p})\}_{\tilde{q},\tilde{p}}&=&
\{A(q,p),B(q,p)\}_{q,p}}
{+{\delta\over \delta q_\ti}\big(\{A,B\}\big)\triangle q_\ti&+&
(q\to p)+O(\triangle q^2,\triangle p^2).}
We would like to stress that when we are performing the second
transformation in (\ref{mw17},\ref{mw18}) which follows the first one, then we
must use the transformed variables. In particular, instead of
$\lam_2(\cn,x,t)$ we must take $\lam_2(\hi_{\lam_1}\cn,x,t)$.
This explains the appearence of the last terms in \refb{mw19}.

First let us consider the case, when the transformations are generated
only by the constraints without extra compensating "trivial" transformations.
In the particular case where the structure coefficents $t^\tal_{\tbe\tga}$
do not depend on the canonical variables $q,p$ the parameter $\bar{\lam}$
also does not depend on them as can be seen from \refb{mw19}.
Also, $\dot{t}^\tal_{\tbe\tga}=0$ in this case and thus the commutator of
two transformations generated by the constraints only ($\xi=0$)
yields again a transformation generated by the constraint.
Hence, {\em if the structure coefficients do not
depend on the canonical variables the transformations generated
by the constraints form a closed algebra off mass-shell.}
On the other hand, if the structure coefficients do depend on the canonical
variables that does not automatically imply that the algebra of
transformations will not close even in the absence of trivial
transformations. Actually, the $q,p$-dependence
in the formula \refb{mw19} for $\bar{\lam}$ can, in principle, be
cancelled. The price we pay for that is that the $\lam$-parameters
may become $\cn$-dependent. As we shall see in section 7 this happens
for gravity where some of the structure coefficients depend
on $q$, if we consider transformations of functions which depend
on the canonical variables \refb{mw17}.
The algebra of transformations \refb{mw10} can also be closed
in all relevant cases when $\xi\neq 0$ if the $\lam^\tbe$
and $\xi$ are related in a certain way. The corresponding
transformations can be interpreted as Lagrangean symmetries
when some of the constraints are nonlinear in the momenta.
We shall discuss the cases which are of particular interest for us
later on.

An interesting question to which we have no general answer is the
following: what are the {\em sufficient} conditions to exponentiate
the infinitesimal transformations \refb{mw10} to finite ones. In the theories
we shall consider we know the finite Lagrangean symmetries
which can be formulated in the Hamiltonian formalism
and this way one can find the finite transformation in the first
order formalism. But in general it seems unlikely that the closing of
the algebra of infinitesimal transformations is sufficient
to exponentiate them since already for a
free {\it nonrelativisic particle}, which very probably
does not admit any finite local symmetry, the transformations
\refb{mw8} form a closed algebra.
This difficult and very important question (i.e. for the functional
integral) what are the conditions such that the transformations
\refb{mw10} can be made finite needs further investigation.

\paragraph{Constraints and the equations of motion.} There is a
very interesting and non-trivial connection between the equations
of motion and the constraints.
As it is wellknown, if we demand that the constraints are fulfilled on
some initial hypersurface $t=t_0$, then due to the equations of motion
they will be satisfied at later times. Actually, we have
\eqngr{\dot{C}_\tal &=&
{\delta C_\tal\over \delta q_\ti}\dot{q}_\ti+
{\delta C_\tal\over \delta p^\ti}\dot{p}^\ti}
{&=&\cn^\tbe t^\tga_{\tal\tbe} C_\tga+t^\tbe_\tal C_\tbe+
{\delta C_\tal\over \delta q_\ti}
EM(q_\ti )+{\delta C_\tal\over \delta p^\ti }EM(p^\ti),}
from which immediately follows that $\dot{C}\sim C$ if the
equations of motion are satisfied. Thus we need to
impose the constraints only on the initial hypersurface and then
they will hold at any moment of time owing to the equations of motion.\par
Inversely, in some theories (e.g. gravity) we can get all of the equations
of motions (or some of them as in string theory) if we only demand that the
constraints are fulfilled for all $t$ (i.e. everywhere) and that the symmetry
transformations do not destroy this property. For example, in gravity
and string theory this means that we demand that the constraints are
valid everywhere and for any choice of spacelike hypersurfaces, because
the symmetry transformations (diffeomorphism transformations) can
be interpreted as a change of foliation of space-time.
In general relativity this statement is known as {\em interconnection
theorem} \cite{i92}. Usually, to prove this theorem it is assumed that the
first four Einstein equation corresponding to the constraints
are valid in any coordinate system (for any foliation) and then one
immediately concludes that this can be true only
if the remaining six Einstein equations are satisfied. Moreover,
if one considers finite transformations then it suffices to
demand that only the first Einstein equation must be fulfilled
to conclude that the remaining equations must hold \cite{i92}.

Note however, that to get the Einstein equations one needs to
impose half of the Hamiltonian equations to express the momenta
in term of the velocities. In the Hamiltonian formulation these
equations are on the same footing as the other ones and thus
the above arguments can hardly be seen as satisfactory in
a Hamiltonian approach since we cannot state that the {\em whole}
dynamics is encoded in the constraints. One does not deduce
the Hamiltonian equations of motion only from the constraints
and symmetries. Thus our purpose will be to fill this gap
and derive the equations of motion using the constraints and the
symmetry properties entirely in the Hamiltonian formalism,
without postulating any of the Hamiltonian equations of motion.

For that let us consider how the constraints are changed under the symmetry
transformations \refb{mw10}:
\eqngrl{
\dxila C_\tal&=&{\delta C_\tal\over \delta q^x_i}\dxila q^x_i
+{\delta C_\tal\over \delta p^{ix}}\dxila p^{ix}}
{&=&{\delta C_\tal\over \delta q^x_i}EM(q^x_i)\xi^x+
  {\delta C_\tal\over \delta p^{ix}}EM(p^{ix})\xi^x
  +\lambda^\tga t^\tbe_{\tal\tga}C_\tbe.}{mw20}
For the known theories the constraints are local functions of
$q$ and $p$ and involve only spatial derivatives of $q$ up to second and $p$
up to first order. It follows then that the
functional derivatives of the constraints have the form
\eqngrl{
{\delta C_{\al y}\over \delta q^x_i}&=&A^i_\al\delta(x,y)+B^{ia}_\al
{\pa\over \pa y^a}\delta(x,y)+D^{iab}_\al {\pa ^2\over \pa y^a \pa y^b}
\delta (x,y)}
{{\delta C_{\al y}\over \delta p^{ix}}&=&E_{\al i}\delta(x,y)+F^a_{\al i}
{\pa\over \pa y^a}\delta(x,y),}{mw21}
where $A,B,\dots$ are functions of $q^y$ and $p^y$. Substituting
\refb{mw20} into \refb{mw21} a straightforward calculation yields
\eqngrrl{
\dxila C_{\al y} &=&\big(\dot{C}_{\al y}+\cn^\tbe t^\tga_{\tbe ,\al y}
C_\tga +t^\tga_{\al y} C_\tga\big)\xi_y}
{&+&\lam^\tga t^\tbe_{\al y,\tga} C_\tbe+\Big(B^{ia}_\al EM(q^y_i)+
F^a_{\al i} EM(p^{iy})\Big){\pa\xi^y\over \pa y^a}}
{&+&D^{iab}_\al\Big(2{\pa EM(q^y_i)\over \pa y^a}{\pa\xi^y\over \pa y^b}
+EM(q^y_i){\pa^2\xi^y\over \pa y^a\pa y^b}\Big).}{mw63}
Here we used the explicit form for some of the indices $\tal=\al,y$,
$\ti=i,x$ etc; $a,b$ run over the spatial indices and it is understood
that there is no integration over $y$.

Now we can reformulate our question in the following manner:
when can the equations of motion (or some of them)
be the consequence of the equations
\begin{equation}
C_\tal=0\quad\hbox{and}\quad\delta_{\xi,\lam}C_\tal=0.\label{mw22}
\end{equation}
The first condition just means that the constraints are fulfilled
everywhere and the second one that this statement does not depend
on the chosen foliation.

{}From \refb{mw20} we can immediately conclude that the equations
of motion can be derived from \refb{mw22} only if the following necessary
conditions are satisfied:
\begin{itemize}
\item Some of the constraints should be nonlinear in the momenta,
since, as we showed earlier, only in this case should we use the
extra "trivial" transformations (and consequently $\xi\neq 0$).
\item The system should have an infinite number of degrees of
freedom. Otherwise there are no spatial derivatives of $\xi$
and the pieces which are proportional to the equations of motion are absent.
\item The constraints should involve spatial derivatives of the
$p$ and/or the $q$. Else all coefficients $B,F,D$ in \refb{mw21} vanish
and the pieces proportional to the equations of motion are again
absent.
\end{itemize}
If we demand that \refb{mw22} holds for an arbitrary $\xi$, then from
(\ref{mw20},\ref{mw21}) we immediately get the following set of equations
\eqngrrl{
D_\al^{iab}EM(q_i^y)=0}
{B_\al^{ia}EM(q_i^y)+2D_\al^{iba}{\pa EM(q^y_i)\ov \pa y^b}=0}
{F_{\al i}^a EM(p^{iy})=0}{mw23}
which can be solved to obtain the equations of motion.
The equations of motion which we can get from \refb{mw23}
depends on the properties of the matrices $D,B,F$.

Now we will briefly review how the general results apply
to particular systems:
\paragraph{Systems with a finite number of degrees of freedom:}
In this case no equations of motion follow from \refb{mw22} even if
$\xi\neq 0$ since there are no spatial derivatives of $\xi$.
\paragraph{Gauge theories} All of the constraints are linear
in the momenta and therefore the "trivial" transformations
\refb{mw7} are absent. Consequently, none of the equations of motion
can be obtained from \refb{mw22}.
\paragraph{Bosonic string:} One constraint
is nonlinear in the momenta and hence $\xi\neq 0$. The matrices
$F,D$ are identically zero and $B\neq 0$.
Then only some relations between the $EM(q)$ follow from
\refb{mw22} (see section 6).
\paragraph{Gravity:}  This is the most
interesting case. One constraint is nonlinear and leads to
$\xi\neq 0$ for the diffeomorphism transformations. The matrices
$F$ and $D$ are non-singular. As
is clear from \refb{mw23} all Hamiltonian equations
follow then from \refb{mw22}, that is the whole dynamics of general
relativity in the Hamiltonian formulation is hidden in the
requirement that the constraints are satisfied everywhere and
for any foliation. Let us stress that in distinction to \cite{i92}
we did not assume $EM(q)=0$. These equations are also consequences
of eqs. \refb{mw22} and thus the interconnection theorem has been
proven entirely within the Hamiltonian formalism (see section 7).

In the following we apply the general results of this
section to concrete systems. First in sect. 3 to gauge theories
which are trivial as regarding their symmetries, since all
constraints are linear in the momenta and thus the gauge
transformations are generated by the constraints themselves.
Then we consider the relativistic particle where the
constraint is quadratic in the momenta. We demonstrate the
role played by the "trivial" transformations to recover
the reparametization invariance. In sect. 5 we show how
to proceed if several constraints are nonlinear in the
momenta at the example of the locally supersymmetric relativistic
particle \cite{bdz76}. The sections 6 and 7 are devoted to string theory and
gravity. The different sections are selfcontained and the
reader may skip those parts which are not of immediate interest
for him/her.

\chapter{Yang Mills-theories}
The action for the (non-abelian) gauge fields is
\begin{equation}
S=-{1\over 4}\int \tr\big[F_{\mu\nu}F^{\mu\nu}\big]d^3x dt
\end{equation}
where \footnote{$a,b,\dots$ denote internal indices, $\mu,\nu\dots$
space-time indices. The $T_a$ are hermitean generators and the
structure constants $\str$ are totally antisymmetric.}
\eqngr{
F_{\mu\nu}&=&\pamu A_\nu-\panu A_\mu -i[A_\mu,A_\nu]}
{A_\mu&=& A^a_\mu T_a,\qquad [T_a,T_c]=i\str T_c,}
and it is invariant under local gauge transformations
\begin{equation}
A_\mu\longrightarrow e^{-i\theta}A_\mu e^{i\theta}+ie^{-i\theta}
\pamu e^{i\theta}\label{mw24}
\end{equation}
with $\theta=\theta^a T_a$. The functions $\theta^a=\theta^a(x,t)$
are arbitrary functions on space-time. The infinitesimal form of
these gauge transformations is
\begin{equation}
\delta_\theta A_\mu^a=-\pamu \theta^a -f^a_{b c}A_\mu^b \theta^c
=-(D_\mu\theta)^a.\label{mw25}
\end{equation}
Now we will show that these infinitesimal gauge transformations are
just the transformations generated by the constraints (see \refb{mw6}).

In the usual way one can now transform the Lagrangean system
into the corresponding Hamiltonian system and obtains
the following first order action for Yang-Mills theories \cite{r81}
\begin{equation}
S=\int \Big[\unpi_\ta\cdot{\dot{\una}} ^\ta-
A^0_\ta\big(\und\cdot\unpi\big)^\ta-\ha
\big(\unpi_\ta\cdot \unpi^\ta+\unb_\ta\cdot\unb^\ta\big)\Big]dt,\label{mw26}
\end{equation}
where $\ta=(a,x)$, $\pi^\ta_i$ are the momenta conjugate
to $A^\ta_i$, and
\eqngr{
\big(\und & \cdot\unpi\big)^a=\unpa\cdot\unpi^a+f^a_{bc}\una^b\cdot \unpi^c}
{\unb^a&=-\unpa\times\una^a-\ha f^a_{bc}\una^b\una^c.}
Here we collected the spatial components into $3$-vectors
$\una =(A_1,A_2,A_3)$ (similarly for $\unpi,\,\unb$) and assume
the gauge group to be compact, so that $\una^a=\una_a$ etc.

The system \refb{mw26} is a first class Hamiltonian system \refb{mw1}
for which the components $A^0_\ta$ play the role of Lagrangean multipliers,
the constraints are just
\begin{equation}
C_\ta =(\und\cdot\unpi)_\ta,\label{mw27}
\end{equation}
and the Hamiltonian
\begin{equation}
\ch=\ha\big(\unpi^\ta \unpi_\ta+\unb^\ta \unb_\ta\big).
\end{equation}
The constraints and Hamiltonian form a closed algebra with
respect to the Poisson bracket
\begin{equation}
\{C_{ax},C_{by}\}=\str\delta(x,y)C_{cx},\qquad
\{\ch,C_{ax}\}=0.
\end{equation}
{}From that it follows that the structure coefficients are equal to
\begin{equation}
t^{cz}_{ax,by}=f^c_{ab}\;\delta (x-y)\delta(z-x),\qquad
t^{ax}_{by}=0.\label{mw28}
\end{equation}
Substituting \refb{mw27} and \refb{mw28} in formulae \refb{mw6} we obtain
the following symmetry transformations for the system \refb{mw26}
\begin{equation}
\delta\una^\ta=\{\una^\ta,\lam^\tb C_\tb\}=-(\und\lam)^\ta\qquad
\delta A^0_\ta=\delta \cn^\ta=\dot{\lam}^\ta-t^\ta_{\tb\tc}A^{0\tb}
\lam^\tc\label{mw29}
\end{equation}
and
\begin{equation}
\delta\unpi_\ta=\{\unpi_\ta,\lam^\tb C_\tb\}=-f^a_{bc}\unpi^{bx}\lam^{cx}.
\label{mw30}
\end{equation}
These transformations correspond to symmetries of the corresponding
Lagrangean system since the constraints \refb{mw27} are linear
in the momenta.
The transformations \refb{mw29} coincide with \refb{mw25} if we identify
$\lam=\theta$. Hence it is clear that the whole group of
gauge transformations (including time dependent ones) is generated
by the constraints. It is easy to verify that the transformations
for the momenta \refb{mw30} follow from the first equation in \refb{mw29}
if we use the relation between velocities $\dot{\una}_\ta$
und momenta $\unpi_\ta$ (the first Hamiltonian equation)
which defines the supspace $\cm$ where the Lagrangean system
lives. To compare the symmetries in the Lagrangean and Hamiltonian
formulations we need to use these equations. However, the Lagrangean
system lives in the subspace $\cm$ (see \refb{mw5}) while
the transformations (\ref{mw29},\ref{mw30}) can be viewed as symmetries
in the whole phase space and hence the group of symmetries
is richer in the Hamiltonian formalism since it acts also on
trajectories which do not belong to $\cm$.

The transformations (\ref{mw29},\ref{mw30}) can be made finite in phase space
off the hypersurface $\cm$. Actually the action \refb{mw26} is invariant
under the transformation \refb{mw24} if simultaneously the momenta
are transformed as
\begin{equation}
\pi\longrightarrow e^{-i\theta}\pi e^{i\theta}.
\end{equation}
To prove this we do not need to use any of the Hamiltonian
equations. So this symmetry holds for all trajectories in
phase space. This is why the global symmetry of Hamiltonian
systems is richer as the usual gauge symmetry of the corresponding
Lagrangean systems.

\chapter{Relativistic particle}
It is convenient to describe the relativistic particle
moving in $4$-dimensional Minkowski spacetime by $4$ scalar fields
$\phi^\mu(t)$, $\mu=0,1,2,3$, in $1$-dimensional 'spacetime'
with coordinate $t$. The action has the form
\begin{equation}
S=-\ha\int\sqrt{-g}\big[g^{00}\dot\phi^\mu\dot\phi_\mu+m^2\big]dt\label{mw31}
\end{equation}
where the dot denotes differentiation with respect to time $t$ and
$\phi^\mu\phi_\mu=-(\phi^0)^2+\sum_1^3 (\phi^i)^2$. The
$m^2$ term maybe viewed as 'cosmological constant' in $1$-dimensional
'spacetime' with metric $g_{00}$.

The action \refb{mw31} is manifestly invariant under general
coordinate transformations in $1$-dimensional 'spacetime'
(reparametrization invariance). The infinitesimal form of these
transformations reads
\begin{equation}
t\to t-\xi,\qquad g_{00}\to g_{00}+\cl_\xi g_{00},\qquad
\phi^\mu\to \phi^\mu+\cl_\xi \phi^\mu,\label{mw32}
\end{equation}
where $\cl_\xi$ is the Lie-derivative in $1$-dimensional 'spacetime'.
Introducing  the lapse funcion $\cn$ according to
\begin{equation}
g_{00}=-\cn^2\label{mw41}
\end{equation}
and defining the momenta conjugated to $\phi_\mu$
\begin{equation}
\pi_\mu={\pa\cl\ov\pa\dot\phi_\mu}={\dot\phi^\mu\ov\cn}\label{mw33}
\end{equation}
as a result of the Legendre transformation one finds the following
first order action for the relativistic particle
\begin{equation}
S=\int\big[\pi^\mu\dot\phi_\mu-\cn C\big]dt.\label{mw34}
\end{equation}
The lapse function $\cn$ plays the role of a Lagrangean multiplier
in \refb{mw34} and the constraint is {\em quadratic in the momenta}
\begin{equation}
C=\ha\big(\pi^\mu \pi_\mu +m^2\big).\label{mw35}
\end{equation}
Of course, the structure coefficient vanishes.

The action \refb{mw34} still should be invariant (at least in $\cm$)
under the infinitesimal diffeomorphisms \refb{mw32}, the explicit
form of which is
\begin{equation}
\delta\phi^\mu=\dot\phi^\mu\xi\quad\hbox{and}\quad
\delta \cn=(\cn\xi)^\cdot\;.\label{mw36}
\end{equation}
Since $\pi$ and $\phi$ are independent variables in the first
order formalism, we should add to \refb{mw36} the transformation
law for $\pi$ to have the diffeomorphisms on the whole phase space.
This transformation law for $\pi$, {\em which corresponds to the
diffeomorphism group}, can be obtained at first in $\cm$ (see \refb{mw5}),
where the Lagrangean system lives, from \refb{mw33} as
\begin{equation}
\delta\pi^\mu={\delta\dot\phi^\mu\ov\cn}-{\dot\phi^\mu \ov\cn^2}
\delta\cn=\dot\pi^\mu\xi\label{mw37}
\end{equation}
and then can be extended to the whole phase space and hence to
trajectories for which \refb{mw33} does not hold. Clearly
the transformations (\ref{mw36},\ref{mw37}) correspond to the
reparametrisation (diffeomorphism) invariance of the relativistic
particle in the Hamiltonian formalism. They coincide
with \refb{mw16}\footnote{For systems
with a finite number of degrees of freedom $\xi^y-\xi^x$ vanishes
in \refb{mw16}}, which is a special combination of the 'trivial' and
constraint-generated transformations.

The algebra of transformations (\ref{mw36},\ref{mw37}) closes and
forms a Lie algebra on the whole phase space. Their finite form
reads
\begin{equation}
\phi^\mu(t)\to \phi^\mu(\tau(t)),\quad
\pi_\mu(t)\to \pi_\mu(\tau(t)),\quad
\cn(t)\to {d\tau\over dt}\cn(\tau(t)).\label{mw38}
\end{equation}
It is easy to see (without using Hamilton's equations) that the
action \refb{mw34} is invariant under these finite tranformations
completely off mass-shell.

The first order action \refb{mw34} is also (off mass-shell) invariant
under the transformations \refb{mw6} generated by the constraints
alone
\begin{equation}
\delta_\lam\phi^\mu=\lam\pi^\mu,\quad
\delta_\lam \pi_\mu=0,\quad
\delta\cn=\dot{\lam}.\label{mw39}
\end{equation}
It is clear that they are very different from the reparametrisation
transformations (\ref{mw36},\ref{mw37}) even in the subspace $\cm$
and hence cannot correspond to any Lagrangean symmetry. Only
{\em on mass-shell},
\begin{equation}
\dot\phi^\mu=\cn\pi^\mu\quad,\quad \dot\pi^\mu=0
\end{equation}
do the transformations \refb{mw39} coincide with the reparametrization
transformations if we make the identification $\lam=\cn\xi$. However, as we
argued earlier the comparison of infinitesimal transformations on
mass-shell is meaningless.

If we demand that as a result of the transformation \refb{mw39}
the trajectory should stay in $\cm$ then we immediately see that
this can be true only for on-shell trajectories. Therefore
\refb{mw39} can be viewed as the dynamical equations in the
subspace $\cm$, where the Lagrangean system lives. Thus we
conclude that the nonlinear constraint \refb{mw35} generates
the dynamics, rather than symmetries in $\cm$. This explains
the origin of the dynamics for superhamiltonian systems.

However, in the whole phase space, the infinitesimal transformations
\refb{mw39} can still be viewed as {\em off mass-shell} symmetries
of the Hamiltonian system. Moreover they can be exponentiated to
the finite ones
\begin{equation}
\phi^\mu(t)\to\phi^\mu(t)+\lam(t)\pi^\mu(t),\quad
\pi_\mu(t)\to \pi_\mu(t),\quad
\cn(t)\to\cn(t)+\dot{\lam}.\label{mw40}
\end{equation}
which are very different from \refb{mw38}. As we stressed
already, the symmetry (\ref{mw39},\ref{mw40}) do not correspond
to diffeomorphisms of the Lagrangean system and it is not clear
to us what is the relevance of this symmetry which exists
only in the Hamiltonian version of the theory.

\chapter{The locally supersymmetric relativistic particle}
The theory of the relativistic particle can be super-symmetrized
and this leads to the simplest one-dimensional analog of supergravity, namely
the theory for the locally supersymmetric relativisic particle \cite{bdz76}.
For that we need to introduce in addition to the bosonic variables
$\phi^\mu$ fermionic variables $\psi^\mu$
which live in 1-dimensional 'spacetime' and they
would describe spin-$1/2$ particles in 4-dimensional
spacetime. To make the theory locally supersymmetric we also need
the analog of the the spin-$3/2$  gravitino field in
supergravity and which we denote by $\chi$. Then the action for massless
particles reads
\begin{equation}
S=-\ha\int dt\det (e_0^{\,\hnn})\big[g^{00}\dot{\phi}^2-i\bar{\psi}\gamma^0
\dot{\psi}-g^{00}\bar{\chi}_0\psi\dot{\phi}\big].\label{mw42}
\end{equation}
To simplify the formulae we skipped all external indices. Here
$e_0^{\,\hnn}$ is the einbein field in 1-dimensional 'spacetime'
on which the bosonic fields $\phi,e_0^{\,\hnn}$ and fermionic ones
$\psi,\chi_0$ live. We denote by $\hnn$ the 'Lorentzian' index and
by $0$ the 'spacetime' index. The fermionic fields are assumed to
be real Majorana fields and the 'spin $3/2$' field $\chi$ is taken
in the Rarita-Schwinger representation where it is considered
as covariant vector of Majorana spinors. Of course in one-dimensional
'spacetime' this covariant vector has only one component.\par
Introducing the lapse function as in \refb{mw41} and taking into account
that
\begin{equation}
e_0^{\,\hnn}=\cn,\quad e^0_{\,\hnn}={1\over \cn},\quad
\gam^0=e^0_{\,\hnn}\gamma^{\hnn}={i\over \cn}
\end{equation}
and for Majorana spinors
\begin{equation}
\bar{\psi}=\psi^\dagger\gam^{\hnn}=i\psi\quad,\quad
\bar{\chi}=\chi^\dagger\gam^{\hnn}=i\chi
\end{equation}
the action \refb{mw42} becomes
\begin{equation}
S=\ha\int dt\big[{1\over \cn}\dot{\phi}^2-i\psi\dot{\psi}-{i\over \cn}
\chi\psi\dot{\phi}\big].\label{mw43}
\end{equation}
This action is manifestly invariant under (infinitesimal) diffeomorphism
transformations
\begin{equation}
t\to t-\xi^0\quad\hbox{and}\quad Q\to Q+\cl_\xi Q
\end{equation}
which now have the explicit form
\begin{equation}
\delta \phi=\dot{\phi}\xi^0,\quad
\delta \psi=\dot{\psi}\xi^0,\quad
\delta \cn=(\cn\xi^0)^\cdot,\quad
\delta \chi=(\chi\xi^0)^\cdot,\label{mw44}
\end{equation}
since $\phi,\psi$ are spacetime scalars and $\chi$ is a covariant
vector. In addition it is also invariant under (infinitesimal)
local supersymmetry transformations
\begin{equation}
\delta\phi=i\theta \psi,\quad
\delta\psi=\theta(\dot{\phi}-{i\over 2}\chi\psi)\cn^{-1},\quad
\delta\cn=i\theta \chi,\quad
\delta\chi=2\dot{\theta},\label{mw45}
\end{equation}
where $\theta$ is the time-dependent Grassmannian parameter of the
supersymmetry transformations. Clearly the action \refb{mw43} is invariant
under simultaneous {\em infinitesimal} diffeomorphisms \refb{mw44} and
supersymmetry transformations \refb{mw45}. Our aim is to
recover the corresponding off mass-shell symmetries (diffeomorphisms
and local supersymmetry) in the first order Hamiltonian formalism.

The standard procedure leads to the following first order action for the
locally supersymmetric particle
\begin{equation}
S=\int \big[\pi_\phi\dot{\phi}-\ha i\psi\dot{\psi}-\cn^\al C_\al]dt
\label{mw46}\end{equation}
with Lagrangean multiplier fields
\begin{equation}
\cn^0=\cn\quad\hbox{and}\quad \cn^1=\ha\chi.
\end{equation}
Thus $\cn^0$ is the bosonic lapse function and $\cn^1$ proportional to the
fermionic 'gravitino' field. The constraints
\begin{equation}
C_0=\ha \pi^2_\phi\qquad\hbox{and}\qquad C_1=i\pi_\phi\psi\label{mw47}
\end{equation}
form a closed algebra with respect to the Poisson bracket, which
are generalized to graded algebras to include fermionic variables
as follows:
\begin{equation}
\{\phi,\pi_\phi\}=1\quad,\quad \{\psi,\psi\}=i.\label{mw48}
\end{equation}
Actually we have
\begin{equation}
\{C_0,C_0\}=\{C_0,C_1\}=0\quad\hbox{and}\quad
\{C_1,C_1\}=-2iC_0.
\end{equation}
As it follows from \refb{mw48} the only nonvanishing structure coefficient
is
\begin{equation}
t^0_{11}=-2i.
\end{equation}
The infinitesimal transformations \refb{mw6} generated
by the constraints \refb{mw47} read
\eqngrl{
\delta_\lam\phi=\lam^0\pi_\phi +i\lam^1\psi,\quad \delta_\lam\pi_\phi &=&0,
\quad \delta_\lam\psi=\lam^1\pi_\phi}
{\delta_\lam\cn=\dot{\lam}^0+i\lam^1 \chi\quad &&\quad
\delta_\lam\chi=2\dot{\lam}^1.}{mw49}
where $\lam^0$ and $\lam^1$ are bosonic and Grassmannian variables,
respectively. Actually they are nilpotent, i.e. $\{\{.,C\}C\}=0$, and
thus can easily be exponentiated to finite ones. One obtains the finite
transformations
$F(t)\to F(t)+\delta_{\lam(t)}F(t)$, where $F(t)$ denotes
any of the fields or Lagrangean multipliers appearing in \refb{mw49}.
However, the transformations \refb{mw49} are not really the symmetries
of the Lagrangean system we are looking for.

To see that more clearly we first write the equations of motion
which are gotten by varying the action \refb{mw48} with respect to the
dynamical variables $\phi,\pi_\phi$ and $\psi$
\eqngrr{
EM(\phi)&=&\dot{\phi}-\cn\pi_\phi-{i\over 2}\chi\psi=0,}
{EM(\pi_\phi)&=&\dot{\pi}_\phi=0,}
{EM(\psi)&=&\dot{\psi}-\ha\pi_\phi\chi=0.}
The subspace $\cm$ in which the Lagrangean system lives is defined
by the eq. $EM(\phi)\es 0$. In this subspace
we can read that the momentum $\pi_\phi$
under the transformations \refb{mw44} and \refb{mw45} should be
tansformed as
\begin{equation}
\delta\pi_\phi=\dot{\pi}_\phi \xi^0 +{i\over \cn}\theta
\big(\dot{\psi}-\ha\pi_\phi\chi\big).\label{mw50}
\end{equation}
Comparing (\ref{mw44},\ref{mw45}) and \refb{mw50} with the transformations
\refb{mw49} we immediately conclude that they coincide only if
{\em all} equations of motion are satisfied, that is on mass-shell.

This agrees with our general considerations in sec.2 since for
the supersymmetric particle both constraints in the
action \refb{mw46} are quadratic in the momenta $\pi_\phi$
and $\pi_\psi=-{i\over 2}\psi$ and hence the off-shell
symmetries which correspond to the symmetries of the Lagrangean
system cannot be generated by the constraints alone.
Both of them take a off-shell trajectory which belongs
to $\cm$ away from this subspace. To return the trajectory back
to $\cm$ we need compensating transformations from the set of
trivial transformations \refb{mw7}, one per nonlinear constraint.
In particular the trivial transformations \refb{mw8} in
combination with the transformations generated by the constraint
$C_0$ lead to the familiar reparametrization invariance \refb{mw16}.

Because the constraint $C_1$ is also quadratic in the momenta
we take an extra compensating transformation from the set \refb{mw7}
and add it to \refb{mw10} to obtain all
Lagrangean symmetries in the Hamiltonian formalism. The
resulting transformations read
\begin{eqnarray}
\delta\phi&=&EM(\phi)\xi+\{\phi,\lam^\al C_\al\}=\big(
\dot{\phi}-\cn \pi_\phi-{i\over 2}\chi\psi\big)\xi+\delta_\lam \pi\nonumber\\
\delta\pi_\phi&=&EM(\pi_\phi)\xi+EM(\pi_\psi)\zeta+\{\pi_\phi,\lam^\al C_\al\}
=\dot{\pi}_\phi-i\big(\dot{\psi}-\ha\pi_\phi\chi\big)\zeta,\nonumber\\
\delta\psi&=&EM(\psi)\xi+EM(\phi)\zeta+\{\psi,\lam^\al C_\al\}\label{mw51}\\
&=&\big(\dot{\psi}-\ha\pi_\phi\chi\big)\xi+\big(\dot{\phi}-
\cn\pi_\phi-{i\over 2}\chi\psi\big)\zeta+\delta_\lam \psi,\nonumber\\
\delta\cn&=&\delta_\lam \cn\nonumber\\
\delta\chi&=&\delta_\lam \chi,\nonumber
\end{eqnarray}
where $\delta_\lam$ is given in \refb{mw49} and $\pi_\psi=-{i\over 2}\psi$.
Here $\zeta$ is the Grassmann parameter of the 'extra' transformation
from the set of transformations \refb{mw7} which we need to correct the
gauge transformation generated by the non-linear constraint $C_1$.
Of course the parameters are not independent and are related
by the requirement that the Hamiltonian symmetry is also
a Lagrangean one. The corresponding condition \refb{mw11}, properly
generalized to include fermionic variables, is satisfied by
\begin{equation}
\xi=\xi^0,\quad \zeta={\theta\over \cn},\quad
\lam^0=\cn \xi^0\quad\hbox{and}\quad \lam^1=\theta+\ha\chi\xi^0,\label{mw52}
\end{equation}
expressing the $4$ parameters $\xi,\zeta,\lam^0,\lam^1$ in
terms of $2$ independent parameters $\xi^0$ and $\theta$.
With this identification the symmetry transformations \refb{mw51}
of the Hamiltonian system are reduced exactly to the original diffeomorphism
and
supersymmetry transformations (\ref{mw44},\ref{mw45}) and \refb{mw50} for the
Lagrangean system {\it without using any of the Hamiltonian
equations}. However, the transformations \refb{mw51} with parameters
\refb{mw52} are symmetries even off the subspace $\cm$.
They also form a closed algebra
on the whole phase space \footnote{It is worth noting that
the transformation \refb{mw51} without the extra $\zeta$ term
form a closed algebra only on mass shell}. Actually, denoting
the transformations \refb{mw51} by $\hi (\xi^0,\theta)$ we find
the following commutator of two subsequent transformations
\begin{equation}
[\hi (\xi^0_2,\theta_2),\hi(\xi^0_1,\theta_1)]=
\hi(\xi^0_3,\theta_3)-\hat{1},\end{equation}
where
\eqngr{
\xi^0_3&=&\dot{\xi}^0_1\xi^0_2-\xi^0_1\dot{\xi}^0_2+{2i\over \cn}
\theta_2\theta_1}
{\theta_3&=&\dot{\theta_1}\xi^0_2-\dot{\theta_2}\xi^0_1+{1\over \cn}
\theta_1\theta_2\chi}
Again this closure holds completely off mass shell.
Hence we expect that the transformations (\ref{mw51},\ref{mw52}) can be
'exponentiated' to finite symmetry transformations on the whole phase
space and thus extend the original group of Lagrangean symmetries.

The same properties we expect to hold for supergravity
theories in more dimensions. But because the computations
are quiet involved we have so far refrained from repeating the above
calculations for these more realistic theories.

We conclude this section by stressing that in the considered supersymmetric
model neither of the constraints generates
a symmetry transformation corresponding to a Lagrangean symmetry.
They rather generate the dynamics of the Lagrangean system in the
Hamiltonian formalism, similar as for the relativistic particle.
\chapter{The bosonic string}
The bosonic string propagating in a $D$-dimensional flat target space
can be viewed as the theory for $D$ massless scalar fields $\phi^\mu,\,
\mu=0,\dots,D-1$ on a $2$-dimensional world-sheet spacetime
with metric $g_{\al\beta}$. The action for this theory can be
written in a manifestly invariant form with respect to diffeomorphism
transformations as \cite{p81}
\begin{equation}
S=-\ha \int \sqrt{-g} g^{\al\beta}{\pa \phi^\mu\over \pa x^\al}
{\pa \phi_\mu\over \pa x^\beta}\,d^2 x,\label{mw53}
\end{equation}
where $x^\al\equiv (t,x)$ are the coordinates in the $2$-dimensional
spacetime. To simplify the formulae we shall skip the target-space
index $\mu$ since it always appears in a trivial way and can easily
be reinserted. \par
The diffeomorphism transformations which are off mass-shell
symmetries of the action \refb{mw53} are
\begin{equation}
x^\al\to x^\al-\xi^\al,\qquad g_{\al\beta}\to g_{\al\beta}+\cl_\xi
g_{\al\beta},\qquad \phi\to \phi+\cl_\xi \phi,\label{mw54}
\end{equation}
where $\xi^\al$ is the infinitesimal parameter.
In addition the action is invariant with respect to Weyl transformations
\begin{equation}
g_{\al\beta}\to \Omega^2(x)g_{\al\beta}\quad\hbox{and}\quad
\phi\to \phi.\label{mw55}
\end{equation}
To arrive at the first order formulation it is convenient to
use the $1+1$-decomposition for the world-sheet metric as \cite{adm62}
\begin{equation}
g_{\al\beta}=-(\cn^2-\cn^1\cn_1)dt^2+2\cn_1 dx dt+\gamma_{11}dx^2,
\end{equation}
where $\cn$ and $\cn_1$ are the lapse and shift functions, respectively.
We rise and lower the spacial index '1' using the metric $\gamma_{11}\equiv
\gamma$ of the $1$-dimensional hypersurface $t$=constant in $2$-dimensional
spacetime. Correspondingly we have
\begin{equation}
\gamma^{11}={1\over\gamma},\quad \cn^1={1\over \gamma}\cn_1,\quad
\sqrt{-g}=\cn\sqrt{\gamma}.\label{mw56}
\end{equation}
Using \refb{mw54} an easy calculation yields the following explicit
transformation laws for
\begin{equation}
\cn^0={\cn\over \sqrt{\gamma}},\label{mw57}
\end{equation}
$\cn^1$ and $\phi$ under diffeomorphism transformations
$x^\al\to x^\al-\xi^\al,\;\,\xi^\al=(\xi^0,\xi^1)$:
\begin{eqnarray}
\delta\cn^0&=&\delta\big({\cn\over \sqrt{\gamma}}\big)=
(\cn^0\xi^0)^\cdot+\cn^{1\pr} (\cn^0\xi^0)-\cn^1(\cn^0\xi^0)^\pr\nonumber\\
&&\qquad +\cn^{0\pr} (\xi^1+\cn^1\xi^0)-\cn^0
(\xi^1+\cn^1\xi^0)^\pr,\nonumber\\
\delta\cn^1&=&(\xi^1+\cn^1\xi^0)^\cdot+\cn^{1\pr} (\xi^1+\cn^1\xi^0)-
\cn^1(\xi^1+\cn^1\xi^0)^\pr\label{mw58}\\
&&\qquad +\cn^{0\pr} (\xi^1+\cn^{1}\xi^0)-\cn^0
(\xi^1+\cn^1\xi^0)^\pr,\nonumber
\\
\delta\phi&=&\dot{\phi}\xi^0+\phi^\pr\xi^1.\nonumber
\end{eqnarray}
Here dot and prime mean the differentiations with respect to
the time and space coordinates $x^0=t$ and $x^1=x$, respectively.
The transformation law for the momentum $\pi$ conjugated to $\phi$,
\begin{equation}
\pi={\pa \cl\over \pa \dot{\phi}}={\sqrt{\gamma}\over \cn}\big(
\dot{\phi}-\cn^1\phi^\pr\big)\label{mw59}
\end{equation}
follows immediately from \refb{mw58} and \refb{mw59}:
\begin{equation}
\delta\pi=\dot{\pi}\xi^0+(\pi\xi^1)^\pr +(\cn^1\pi+\cn^0\phi^\pr )\xi^{0\pr}.
\end{equation}
In the first order Hamiltonian formulation the action \refb{mw53} takes
the form
\begin{equation}
S=\int \big(\pi\dot{\phi}-\cn^\al C_\al\big)dx dt,
\end{equation}
where the Lagrangean multipliers $\cn^\al$ are just the
functions defined in (\ref{mw56},\ref{mw57}) (that is they are
the lapse and shift functions up to $\sqrt{\gamma}$). The constraints
\begin{equation}
C_0=\ha(\pi^2+\phi^{\pr 2}),\qquad\hbox{and}\qquad C_1=\pi\phi^\pr
\end{equation}
form a closed algebra, i.e. are first class constraints, with
respect to the standard Poisson brackets $\{\phi(x),\pi(y)\}=\delta(x,y)$:
\eqngrl{
\{C_i(x),C_i(y)\}&=&C_1(x){\pa\over\pa x}\delta(x,y)
-C_1(y){\pa\over \pa y}\delta(x,y)}
{\{C_0(x),C_1(y)\}&=&C_0(x){\pa\over\pa x}\delta(x,y)
-C_0(y){\pa\over \pa y}\delta(x,y),}{mw62}
where $i=1,2$. Rewriting these relations in terms of the light-cone
constraints $C_0\pm C_1$ we immediately recognize them as Virasoro algebra
\cite{gsw87}.\par
Concerning the symmetries we first
note that the Weyl symmetry \refb{mw55} takes the trivial form
in the Hamiltonian formalism
\begin{equation}
\cn^0= {\cn\over \sqrt{\gamma}}\to
{\Omega\cn\over \Omega \sqrt{\gamma}}=\cn^0,\qquad
\cn^1= {\cn_1\over \sqrt{\gamma}}\to \cn^1,
\end{equation}
so that all variables in the first order action are Weyl invariant.\par
Because one of the constraints, namely $C_0$, is quadratic
in the momentum, we need to combine gauge and reparametrization
transformations as in \refb{mw10} to recover the diffeomorphism invariance
(\ref{mw58},\ref{mw59}) in the Hamiltonian formalism. For the bosonic string
the
explicit transformation \refb{mw10} reads
\begin{eqnarray}
\delta\cn^0&=&\dot{\lam}^0+\cn^{1\pr}\lam^0-\cn^1\lam^{0\pr}+\cn^{0\pr}\lam^1
-\cn^0\lam^{1\pr}\nonumber\\
\delta\cn^1&=&\dot{\lam}^1+\cn^{1\pr}\lam^1-\cn^1\lam^{1\pr}+\cn^{0\pr}\lam^0
-\cn^0\lam^{0\pr}\label{mw60}\\
\delta \phi&=&EM(\phi)\xi+\{\phi,\lam^\tal C_\tal\}
=\big(\dot{\phi}-\cn^0\pi-\cn^1\phi^\pr\big)\xi
+\pi \lam^0+\phi^\pr\lam^1,\nonumber\\
\delta \pi&=&EM(\pi)\xi+\{\pi,\lam^\tal C_\tal\}
=\big(\dot{\pi}-(\cn^0\phi^\pr +\cn^1\pi)^\pr\big)\xi
+(\phi^\pr \lam^0)^\pr+(\pi\lam^1)^\pr,\nonumber
\end{eqnarray}
where we need to assume that the parameters are related by the condition
\refb{mw14}. This condition is solved if we express the parameters
$\xi,\lam^0,\lam^1$ in terms of two independent parameters as
\begin{equation}
\xi=\xi^0,\qquad \lam^0=\cn^0\xi^0={\cn\over\sqrt{\gamma}}\xi^0,\qquad
\lam^1=\xi^1+\cn^1\xi^0,\label{mw61}
\end{equation}
and then we immediately recognize the transformations \refb{mw60}
as diffeomorphism transformations (\ref{mw58},\ref{mw59}) without using the
Hamiltonian equations. Once
again we emphasize that the transformations \refb{mw60} are infinitesimal
symmetry transformations on the whole phase space whereas the
transformations (\ref{mw58},\ref{mw59}) are applicable only to trajectories
on the hypersurface $\cm$.\par
As a first step towards exponentiating the infinitesimal transformations
\refb{mw60}, i.e. make them finite, we should check their algebra. Using
the formulae for the particular choice \refb{mw61} of parameters
it easy to find that the commutator of two subsequent transformations
$\hi_{\xi,\lam}\equiv \hi(\vec{\xi})$, where $\vec{\xi}=(\xi^0,\xi^1)$
becomes
\begin{equation}
[\hi (\vec{\eta}),\hi(\vec{\xi})]=\hi(\cl_{\vec{\eta}}\vec{\xi})-\hat{1},
\end{equation}
completely off mass shell. Hence the algebra of transformations
\refb{mw60} forms a (infinite dimensional) Lie-algebra even off the
subspace $\cm$.

Let us stress once more that the infinitesimal
gauge transformations generated by the constraints only (that is the
transformations \refb{mw60} with $\xi$ set to zero) are not symmetry
transformations which could correspond to the diffeomorpisms
of the Lagrangean system. The nonlinear constraint $C_0$ is reponsible
for the dynamics.\par
The last remark concerns the connection between the constraints
and the equations of motion for the string theory. Calculating
the first functional derivative of the constraints with
respect to the canonical variables we see that the
$B$ and $E$ coefficients in \refb{mw21} are
\begin{equation}
B_0=E_1=\phi^\pr_y\quad,\quad B_1=E_0=\pi_y,
\end{equation}
while the $D$ and $F$ coefficients vanish.
Then the eqs. \refb{mw23} reduce to
\begin{equation}
\phi^{\prime\mu}EM(\phi_\mu)=0\mtxt{and}\pi^\mu EM(\phi_\mu)=0
\end{equation}
where $\mu$ is the target-space index. From these equations we cannot
conclude that all eqs. of motion should be satisfied. However, they
put certain restrictions on the allowed $EM(\phi)$. Since the
coefficients $F$ are equal zero (the constraints do not involve
any spatial derivatives of the momenta) the requirement that the
constraints are satisfied everywhere and for any foliation does not
tell us anything about the eqs. of motion $EM(\pi)\es 0$. We will
see in the next section that the interconnection theorem, which we
just discussed, has much more interesting content in gravity.
\chapter{Gravity}
General relativity without matter has the action
\begin{equation}
S=\int R\sqrt{-g}d^4x\label{mw64}
\end{equation}
(we adapt the sign and units conventions in \cite{dw67})
and is invariant with respect to coordinate (or diffeomorphism)
transformations, the infinitesimal form of which read
\begin{equation}
x^\al\to x^\al-\xi^\al,\qquad g_{\al\beta}\to g_{\al\beta}+
\cl_\xi g_{\al\beta}.\label{mw65}
\end{equation}
Rewriting the metric $g_{\al\beta}$ in the $3+1$-split form \cite{adm62}
\begin{equation}
ds^2=-(\cn^2-\cn_i\cn^i)dt^2+2\cn_i dx^i dt+\gamma_{ij} dx^i dx^j,
\end{equation}
where $\cn$ is the lapse function, $\cn_i$ are the shift functions,
$\cn_i=\gamma_{ij}\cn^j$, and $\gamma_{ij}$ is the metric
of the $3$-dimensional hypersurface $\Sigma_t$ of constant time $t$,
we derive from \refb{mw65} the following explicit transformations for $\cn$,
$\cn^i$,
and $\gamma_{ij}$:
\begin{eqnarray}
\delta \cn&=&(\cn\xi^0)^\cdot-\cn^i(\cn\xi^0),_i+\cn,_m(\xi^m+\cn^m\xi^0)\;,
\nonumber\\
\delta \cn^i&=&(\xi^i+\cn^i \xi^0)^\cdot-(\xi^i+\cn^i\xi^0),_m\cn^m
+\cn^i,_k(\xi^k+\cn^k\xi^0)\nonumber\\
&&\qquad -\cn\gamma^{ij}(\cn\xi^0),_j+\gamma^{ij}\cn,_j(\cn\xi^0)\;,
\label{mw66}\\
\delta \gamma_{ij}&=&(\dot{\gamma}_{ij}-\cn_{i\vert j}-\cn_{j\vert i})
\xi^0+^{(3)}\cl_{{\bf\xi}+{\bf\cn}\xi^0}\,\gamma_{ij}.\nonumber
\end{eqnarray}
Here the comma denotes ordinary differentiation with respect
to the corresponding space coordinate, the bar denotes
covariant derivative in the $3$ dimensional space $\Sigma_t$
with metric $\gamma_{ij}$, $\gamma^{ij}$ is the inverse $3$-dimensional
metric on $\Sigma_t$ and $^{(3)}\cl$ is the Lie derivative
in $\Sigma_t$. This Lie derivative is to be taken in the
direction ${\bf\xi}+{\bf\cn}\xi^0\equiv \{\xi^i+\cn^i\xi^0\}$.\par
In the first order Hamiltonian formalism the $ADM$ action for
pure gravity takes the form
\begin{equation}
S=\int\big(\pi^{ij}\dot{\gamma}_{ij}-\cn^\al\ch_\al\big)d^3x dt,
\end{equation}
where $\pi^{ij}$ are the momenta conjugated to $\gamma_{ij}$ and
the four Lagrangean multipiers are
\begin{equation}
\cn^0=\cn,\qquad\hbox{and}\qquad
\cn^i=\gamma^{ij}\cn_j,
\end{equation}
that is the lapse and shift functions. Correspondingly
the constraints $\ch_\al$ are\footnote{in this section we denote
the constraints by $\ch_\al$, a notation which is widely used
in gravity} \cite{adm62,dw67}
\begin{equation}
\ch_0=G_{ijkl}\pi^{ij}\pi^{kl}-\sqrt{\gam}\;^{(3)}R,\qquad
\ch_i=-2\gam_{ij}\pi^{jl}_{\vert l},\label{mw67}
\end{equation}
where
\begin{equation}
G_{ijkl}={1\over 2\sqrt{\gam}}\big(\gam_{ik}\gam_{jl}+\gam_{il}\gam_{jk}
-\gam_{ij}\gam_{kl}\big),\qquad \gam=\det(\gam_{ij})
\end{equation}
is the metric in superspace \cite{dw67} and
$^{(3)}R$ the instrinsic curvature of the hypersurface
$\Sigma_t$ of constant time $t$. With the help of the fundamental
Poisson brackets
\begin{equation}
\{\gam_{ij}(x),\pi^{kl}(y)\}=
\delta^{(k}_i\delta^{l)}_j\;\delta(x,y)=
\ha\big(\delta^k_i\delta^l_j+\delta^l_i\delta^k_j\big)\delta(x,y)
\end{equation}
one checks that the constraints \refb{mw67} are first class \cite{dw67}
\begin{eqnarray}
\{\ch_0(x),\ch_0(y)\}&=&
\gam^{ij}(x)\ch_j(x){\pa\over \pa x^i}\delta(x,y)
-\gam^{ij}(y)\ch_j(y){\pa\over \pa y^i}\delta(x,y)\nonumber\\
\{\ch_i(x),\ch_0(y)\}&=&\ch_0(x){\pa\over\pa x^i}\delta(x,y)\\
\{\ch_i(x),\ch_j(y)\}&=&
\ch_j(x){\pa\over \pa x^i}\delta(x,y)
-\ch_i(y){\pa\over \pa y^j}\delta(x,y).\nonumber\\
\end{eqnarray}
Let us note that if we add matter (covariantly coupled to
gravity) to \refb{mw64} then the constraints contain extra pieces,
but their algebra remains unchanged. Another interesing
observation is the following: If we use $\sqrt{\gam}\ch_0$
instead of $\ch_0$ as a constraint then the algebra of
constraints looks very much like a natural generalization of the
Virasoro algebra \refb{mw62} to four dimensions.
It is a nontrivial problem where the diffeomorphism invariance
of the original action \refb{mw64} is hidden in the first order
Hamiltonian reformulation of gravity. There have been various attempts
to reveal this symmetry (see, for instance \cite{bk72,ik85,lw90})\par
Three of the constraints, namely the $\ch_i$, are linear in
momenta, so they should generate transformations
which coincide with diffeomorphism transformations. This
has been realized for time independent transformations
some time ago \cite{h58}. However, the fourth
constraint, namely $\ch_0$, is quadratic in the momenta
and hence cannot generate a symmetry of the corresponding
Lagrangean system according to our
general results in section 2. Only combined with a
compensating transformation does it generate the symmetry we
are looking for. Since the Hamiltonian is zero, this
symmetry is exactly the reparametrization invariance \refb{mw16}.
Assuming that the parameters in \refb{mw10} are related such
that the condition \refb{mw14} is satisfied, we can write this
off shell symmetry transformation for gravity in the
following explicit manner
\begin{eqnarray}
\delta\cn&=&\dot{\lam}^0-\cn^j\lam^0,_j+\cn,_j\lam^j,\nonumber\\
\delta\cn^i&=&\dot{\lam}^i-\cn^j\lam^i,_j+\cn^i,_j\lam^j
-\cn\gam^{ij}\lam^0,_j+\gam^{ij}\cn,_j\lam^0,\nonumber\\
\delta\gam_{ij}&=&EM(\gam_{ij})\xi+\{\gam_{ij},\lam^\tal\ch_\tal\}
\label{mw68}\\
&=&EM(\gam_{ij})\xi+{1\over \sqrt{\gam}}\big(2\pi_{ij}-\gam_{ij}\pi\big)
\lam^0+^{(3)}\cl_{\bf \lam}\gam_{ij}\nonumber
\end{eqnarray}
and
\begin{equation}
\delta \pi^{ij}=EM(\pi^{ij})\xi+\{\pi^{ij},\lam^\tal\ch_\tal\}.\label{mw69}
\end{equation}
Here the $5$ parameters $\xi,\lam^\al$ are to be expressed in terms
of the four independent parameters $\xi^\al$ as
\begin{equation}
\xi=\xi^0,\qquad \lam^0=\cn\xi^0,\qquad \lam^i=\xi^i+\cn^i\xi^0\label{mw70}
\end{equation}
to resolve \refb{mw13} and then it becomes evident that \refb{mw68} is
identical to \refb{mw66}. Again
we need not use any of the Hamiltonian equations. A rather
lengthy calculation shows that the transformation law for the
momenta one gets by using the definition of the momenta
in terms of $\gam_{ij},\,\cn_k$ and \refb{mw66} coincides with
\refb{mw69} also off mass shell.\par
Thus we found that in gravity the three constraints which are linear
in the momenta generate the diffeomorphism transformations while
the forth constraint $\ch_0$ does it only in a particular combination
with the 'trivial' transformation \refb{mw8}. This nonlinear in
momenta constraint itself is responsible for the origin of the
dynamics in the subspace $\cm$ in the superhamiltonian reformulation
of gravity.

The important question is how to read off the Lie algebra structure
of the diffeomorphism group in the Hamiltonian formulation.
Because for gravity the structure coefficients depend on the canonical
variables (in distinction from the previous cases) one might
expect that the algebra of infinitesimal transformations
(\ref{mw68}-\ref{mw70})
cannot close in this case. Actually naively the dependence on
the canonical variables can enter in the parameter $\bar\lam$
for the commutator of two infinitesimal transformations with parameters
through the
$\gam$-dependence of the structure coefficients (see \refb{mw19}).
Fortunately, this expectation is not confirmed. In particular, in the formula
\refb{mw19} for the $\bar{\lam}$-parameter this $\gam$-dependence
of the various terms on the right hand side cancels for the
concrete choice \refb{mw70} for the $\cn$-dependence of the parameters
$\lam$. The price we pay for that is the explicit dependence of the
parameters of transformations on the Lagrangean mulitpliers, but
not on the canonical variables $\gam,\pi$.
Starting from the general formulae (\ref{mw17}-\ref{mw19})
a straightforward but rather lengthy calculation shows
that the transformations (\ref{mw68}-\ref{mw70}) form a Lie algebra
completely off mass shell:
\begin{equation}
[\hi (\eta),\hi(\xi)]=\hi (\cl_\xi \eta)-\hat{1},\qquad
\xi=(\xi^0,\dots,\xi^3),\quad
\eta=(\eta^0,\dots,\eta^3),\label{mw71}
\end{equation}
where $\xi^0,\xi^i$ and $\eta^0,\eta^i$ are defined in \refb{mw70},
as it should be for diffeomorphisms.
The formula \refb{mw71} holds even off the hypersurface $\cm$
where the Lagrangean system lives.

There is a deep connection between the constraints and equations
of motion in gravity. Calculating the derivative of the constraints
in this case we shall find that all of the coefficients $A,\cdots,F$
in \refb{mw21} do not vanish.

In particular, taking into account that the index $k$ in the
formulae (\ref{mw21},\ref{mw63}) in the case of gravity is a
composite one,, $i\equiv(j,k);\;a,b$ run over the same
spatial index $l$ and calculating the derivatives of
$\ch_i$ with respect to $\pi^{jk}$ and $\ch_0$ with
respect ot $\gam_{np}$ we find
\begin{equation}
F^l_{ijk}=-2\gam_{i(j}\delta^l_{k)}\mtxt{and}
D_0^{nplk}=-G^{nplk},
\end{equation}
where $G^{nplk}$ is the inverse of the superspace-metric,
$G^{nplk}G_{lkij}=\delta^{(n}_i\delta^{p)}_j$. Then the first
and last equations in \refb{mw23} take the form
\begin{equation}
G^{nplk}EM(\gam_{np})=0\mtxt{and}2\gam_{ij}EM(\pi^{jl})=0.\label{mw72}
\end{equation}
Since the determinants $\det\,G$ and $\det\,\gam$ are not
equal zero the eqs.\refb{mw72} have the unique solution
\begin{equation}
EM(\gam_{np})=0\mtxt{and}EM(\pi^{jl})=0.
\end{equation}
The remaining equations in \refb{mw23} are then automatically fulfilled.
Thus, we see that in general relativity the whole dynamics follows
from the requirement that the constraints are satisfied everywhere
and they are preserved under diffeomorphisms.
\chapter{Discussion}
In the previous sections we revealed the relevant local
symmetries of Lagrangean systems in the first order Hamiltonian
systems. We have seen that all symmetries have a similar structure
in the Hamiltonian approach although they may look quite
differently in the Lagrangean one. All fundamental field theories
in physics, and in particular the ones we considered, are systems
with first class constraints. If the constraints are linear in the
momenta (as in Yang-Mills theories) then they generate the well
known gauge symmetry. If some of the constraints are nonlinear in the momenta
then only very special
combinations of the transformations generated by the constraints
and simple compensating transformations proportional
to the equations of motion correspond to the off mass shell
symmetries of the corresponding Lagrangean system. If only one of the
constraints is nonlinear in the momenta, as in string theory and
gravity, then the symmetries of the system consist of the
gauge transformations generated by the linear constraints
plus an extra reparametrization transformation related to the
nonlinear constraint, but not just generated by this constraint.
This takes place only if the Hamiltonian is equal to zero, i.e.
is a super-Hamiltonian. All wellknown theories with nonlinear constraints
possess a super-Hamiltonian. However, presently we do not know if
there is a deep connection between the non-linearity of some of the
constraints and the super-Hamiltonian character of the system.
If there are more then one nonlinear constraint then one
has to use extra transformations from the huge set of
transformations \refb{mw11} in a combination with the transformations
generated by the nonlinear constraints to recover the
Lagrangean symmetries.

In any case, the wellknown symmetries
of the Lagrangean systems are manifest in the Hamiltonian formalism
and even more transparent there. Different symmetries may look quite
differently in the Lagrangean formalism (for example, local supersymmetry
and diffeomorphisms) but they have the same formal structure
in the Hamiltonian approach. In addition, the symmetry transformations
for the Hamiltonian systems are richer as the corresponding ones
for the Lagrangean systems. This is so since in the Hamiltonian
approach the transformations are acting on the whole phase space
and are symmetries for all off mass shell trajectories. For such
general trajectories the momenta need not be related to the velocities,
as it should be for Lagrangean systems.\par
We considered mainly the infinitesimal form of the symmetry transformations
and checked the algebra of two subsequent infinitesimal transformations.
We found that for all theories we studied (gauge theories, point particle,
bosonic string and gravity) the algebras are closed completely
off mass shell in the whole phase space, even off the subspace
$\cm$ in which the Lagrangean system lives.
In particular, for gravity,
where the structure constants depend on the canonical variables,
we revealed a closed Lie algebra in the Hamiltonian formalism.

Different theories which are invariant under diffeomorphism
transformations (as for example string theory and gravity) have
similar constraint algebras but the constraints look quite
differently. However, for the phase space transformations
belonging to diffeomorphisms to form a Lie algebra the constraints
themselves should have some underlying common structure which
we did not reveal. For example, we could ask what kind of
general conditions the constraints in string theory,
dilaton coupled 2-dimensional gravity, gravity or higher derivative
gravity, the constraints of which are looking quite differently,
should satisfy to close the algebra.
These interesting questions deserve further investigations.\par
The other question concerns the role of the transformations
generated by the nonlinear constraints alone. We showed
that they are responsible for the dynamics of Lagrangean systems
in the superhamiltonianin formalism.

Also we have seen that there is a deep connection between the structure
of the constraints and the dynamics. For example, in string theory
some of the Hamiltonion equations and in gravity all of them
automatically follow if we demand that the constraints are
satisfied everywhere for any foliation of space time. The presence
of the spatial derivatives of $q$ and/or $p$ is responsible
for that on the technical level.\par
One possible application of the developed approach to phase space
symmetries is a way to construct new theories possessing local
symmetries in the Hamiltonian formalism (see, for instance
\cite{frr92,bdz76}). Actually in
many cases the constraints have a clear physical interpretation (as
the Gauss constraints in electromagnetism). So one starts by
introducing contraints in the theory to satisfy some physical
requirements. Then one should commute the constraints (leading
to secondary constraints) such that
the systems of original constraints together with the
secondary ones form a first class system. Note that only
first class constraints generate local symmetries \footnote{For example,
a system with $2n$ second class constraints
can locally be transformed into a system with $n$ first class constraints
and $n$ gauge fixings by a canonical transformation. Thus the
gauge transformations generated by the first class constraints
are automaticlly fixed by the $n$ gauge fixings and no
symmetries survive.}.
The number of constraints is equal to the number of parameters
of the symmetry transformations of the corresponding Lagrangean
system. \par
Another interesting application of the considered formalism
one could find in the quantized theories, in which we are
ultimately interested. For example, in the functional integral
approach to quantum theories it is more natural to consider
the Hamiltonian functional integral as compared to the Lagrangean
one. This is true in particular for theories which are
invariant under diffeomorphisms. In the phase space functional
integral at least the $q,p$-piece of the measure is just the
well-defined Liouville measure.
After performing the integration over the momenta we arrive
at the functional integral in the Lagrangean formulation. However,
even in the simple case of a first order action \refb{mw1} which
is quadratic in the momenta, a $q$-dependent function multiplying
$p^2$ appears in the measure for the Lagrangean functional integral.
For systems where the action is not quadratic in the momenta
or even for gravity the question about the correct measure becomes
quite nontrivial. Also, in the Hamiltonian version of the
BRST-quantization it is not clear which symmetries
(the ones generated by the constraints alone or the symmetries of the
Lagrangean system) should we use to construct the BRST charge for
systems with nonlinear constraints and whether these different
charges lead to the same final quantization.
Only in the simple cases of the relativistic particle and
supersymmetric particle it has been demonstrated that the results in
both cases are the same \cite{npw88}. For both systems the two
kinds of transformations can be written down in finite form.

For field theories this question has not been investigated.
For theories with nonlinear constraints, and in particular gravity,
there are two different BRST charges.
One belonging to diffeomorphisms and one
to the transformations generated by the constraints.
They coincide only if we impose the equations of motion
and this may be the reason why the Batalin-Vilkovisky
theorem \cite{bv77} might break down when the two
compared gauges are not infinitesimally close to each other
\cite{gr92}. The transformations generated by the two BRST charges
differ by trivial transformations. The relevance of these trivial
transformations can already be seen on the perturbative level in
theories with nonclosing algebras (for instance in supergravity \cite{k78}).

One would like to hope that the results obtained in this paper
could help to fill the gap in the study of symmetries of constraint
Hamiltonian systems which, from our point of view, still exist
even on the classical level in the current literature.
\subsubsection{Acknowledgements:}
This work has been supported by the Swiss National Science Foundation.
We wish to thank K. Kiefer, R. Kallosh, C. Isham, P. Hajicek,
J. Halliwell and J. York for helpful discussions.

\end{document}